\documentclass[prb,amssymb,epsfig,showpacs]{revtex4}

 \usepackage{graphicx}
 \usepackage{amsmath}
 \usepackage[bf,small]{caption2}
 \usepackage{floatflt}

\draft

\begin{document}
\title{Propagation of wave packets in randomly stratified media. }
\author{V. D. Freilikher}
\email{freiliv@mail.biu.ac.il} \affiliation{The Jack and Pearl Resnik
Institute of Advanced Technology, Department of Physics, Bar-Ilan
University, Ramat-Gan 52900, Israel}

\author{Yu. V. Tarasov}
\email{yutarasov@ire.kharkov.ua} \affiliation{Institute for
Radiophysics \& Electronics, National Academy of Sciences of Ukraine,
 12 Acad. Proskura St., Kharkov 61085, Ukraine}

\begin{abstract}
The propagation of a narrow-band signal radiated by a point source in a
randomly layered absorbing medium is studied asymptotically in the
weak-scattering limit. It is shown that in a disordered stratified medium
that is homogeneous on average a pulse is channelled along the layers in a
narrow strip in the vicinity of the source. The space-time distribution of
the pulse energy is calculated. Far from the source, the shape of wave
packets is universal and independent of the frequency spectrum of the
radiated signal. Strong localization effects manifest themselves also as a
low-decaying tail of the pulse and a strong time delay in the direction of
stratification. The frequency-momentum correlation function in a
one-dimensional random medium is calculated.
\end{abstract}


\maketitle


\section{Introduction}


\bigskip

The propagation of quantum wave packets and pulses of electromagnetic
radiation in disordered media is a classical problem with a long-standing
history. The continued interest of physicists in this problem is stimulated
both by the quest to better understand such fundamental problems of disorder
as correlations in momentum-energy space, localization of time-dependent
fields, Wigner time delay, etc., and also by the growing number of
applications that pulsed signals find in modern electronics,
telecommunications, optics, and geophysics. Considerable theoretical and
experimental investigations have been expended to study the propagation of
pulses in randomly inhomogeneous media in diffusive regime (see, for
example, Ref.~\onlinecite{Isimaru} and references therein). Much less
studied is the space-time evolution of wave packets in disordered
one-dimensional and layered systems where the interference of multiple
scattered fields is of crucial importance.

It was shown in Refs.~\onlinecite{FreiGred91,6} that in a homogeneous
on average, randomly layered medium where the refractive index
(potential) is a random function of one coordinate only waves
(quantum particles) are localized in the direction of stratification
and propagate along layers forming the so called fluctuational
waveguide. The statistics of wave fields radiated by a monochromatic
point-like source in a randomly layered medium was studied in
Refs.~\onlinecite{6,IEEE}. For its analysis, the resonance expansion
method was applied to calculate correlation functions of plane
harmonics with different ``transverse energies'', i.e. squared
projections of the wave vector on the axis of stratification. In the
case of a non-stationary signal, the problem becomes much more
complicated because it involves correlation analysis of waves with
different both frequencies and transverse wave numbers.

In the present paper, we investigate analytically the space-time
distribution of the average intensity of pulse field radiated by a point
source in a randomly stratified weakly scattering medium with dissipation.
As an intermediate result, the correlation function of the propagators
(Green functions) with different frequencies and transversal wave numbers is
calculated. Localization of the constituent plane harmonics is shown to
result in channelling of the pulse within the fluctuation waveguide and in a
significant modification of the spectral content of the signal far away from
the source. The shape (envelop) of the pulse in the far zone is calculated.
It is shown to be universal and independent on the spectrum of the radiated
packet. This is due to both the filtration of the harmonics by their
localization radii transverse to the layers and to the difference in phase
velocities of those harmonics in this direction. The same reasons cause
strong time delay of the pulse when the receiver is shifted towards the
direction of stratification from the horizontal plane in which the source is
located. This effect is a clear manifestation of the delay time concept
introduced earlier \cite{Wigner} on the basis of scattering phases of
quantum particles moving in disordered media.


\section{Formulation of the problem}


We consider the wave equation for the scalar non-monochromatic field $G({\bf %
R},{\bf R_{0}}|t)$ radiated by a source located at a point ${\bf R_{0}}$ in
an infinite medium which is randomly stratified along $z$-axis,

\begin{equation}
\left[ \Delta -\frac{1}{c^{2}}\frac{\partial }{\partial t}\left( \varepsilon
(z)\frac{\partial }{\partial t}+4\pi \sigma \right) \right] G({\bf R},{\bf %
R_{0}}|t)=4\pi \delta ({\bf R}-{\bf R_{0}})A(t)e^{-i{\omega _{0}}t}.
\label{1}
\end{equation}
Here $\Delta $ is Laplacian, $\varepsilon (z)=\varepsilon _{0}+\delta
\varepsilon (z)$ is the (random) dielectric permeability with the mean value
$\varepsilon _{0}$, $\sigma $ is the conductivity of the medium, $A(t)$ is
the envelope of a wave packet (pulse) with the carrier frequency $\omega
_{0} $. In what follows we consider a narrow-band wave packet which means
that $A(t)$ is a smooth function as compared to the oscillating exponential
in the rhs of equation (\ref{1}).

Since permeability $\varepsilon $ in Eq.~(\ref{1}) depends on one coordinate
only, the problem of finding the mean intensity $\left\langle I\right\rangle
=\big<{|G|}^{2}\big>$ at a given point ${\bf R}$ is reduced, after Fourier
transformation in $\left\{ x,y\right\} $ plane, to calculation of the
one-dimensional correlator of harmonics propagating along the axis of
stratification,

\begin{eqnarray}
\left\langle I\left( {\bf R},{\bf R_{0}}|t\right) \right\rangle &=&\frac{1}{%
\left( 2\pi \right) ^{4}}\int \!\!\!\!\int_{0}^{\infty }d\kappa
_{1}\,d\kappa _{2}\,\kappa _{1}\kappa _{2}\,J_{0}\left( \kappa _{1}\varrho
\right) J_{0}\left( \kappa _{2}\varrho \right)  \nonumber \\
&&\times \int \!\!\!\!\int_{-\infty }^{\infty }d\omega \,d\Omega e^{-i\Omega
t}\varphi \left( \omega +\Omega -\omega _{0}\right) \varphi ^{\ast }\left(
\omega -\omega _{0}\right) K\left( z,z_{0}|\kappa _{1},\kappa _{2};\omega
,\Omega \right) \ .  \label{2}
\end{eqnarray}
Here, angular brackets denote statistical averaging over the ensemble of
random functions $\delta\varepsilon (z)$, $J_{0}\left( x\right) $ is the
Bessel function, $\varrho =\big|{\bf R}_{\parallel }-{\bf R}_{0\parallel }%
\big|$ is the in-plane distance from the source, ${\bf R}_{\parallel }$ is
the radius-vector component parallel to the layers, $\varphi (\omega )$ is
the spectral function of the radiated pulse. Function $K(\ldots )$ in Eq.~(%
\ref{2}) is the two-point correlation function

\begin{equation}
K\left( z,z_{0}|\kappa _{1},\kappa _{2};\omega _{1},\omega _{2}\right)
=\left\langle {\cal G}\left( z,z_{0}|\kappa _{1},\omega _{1}\right) {\cal G}%
^{\ast }\left( z,z_{0}|\kappa _{2},\omega _{2}\right) \right\rangle
\label{3a}
\end{equation}
that we present in a form more convenient for subsequent calculation by
changing the integration variables, viz.

\begin{equation}
K\left( z,z_{0}|\kappa _{1},\kappa _{2};\omega ,\Omega \right) =\left\langle
{\cal G}\left( z,z_{0}|q^{2}+\Delta q^{2},\omega +\Omega \right) {\cal G}
^{\ast }\left(z,z_{0}|q^{2},\omega\right) \right\rangle \ .  \label{3}
\end{equation}
Here $\omega =\omega_{2}$, $\Omega=\omega _{1}-\omega_2$, $q^{2}=\kappa
_{m}^{2}-\kappa_2 ^{2}$, $\kappa_{m}^{2}=\varepsilon _{0}\left( \omega
/c\right) ^{2}$. The ``energy'' difference $\Delta q^{2}$ is given by

\begin{equation}
\Delta q^{2}=\kappa _{2}^{2}-\kappa _{1}^{2}+\varepsilon _{0}\frac{(\omega
+\Omega )^{2}-\omega ^{2}}{c^{2}}+i\left[ \gamma \left( \omega +\Omega
\right) {\left( \frac{\omega +\Omega }{c}\right) }^{2}+\gamma (\omega
)\left( \frac{\omega }{c}\right) ^{2}\right] \ ,  \label{5}
\end{equation}
$\gamma (\omega )=4\pi {\sigma }/{\omega }$. In Eq.~(\ref{3}), ${\cal G}%
\left( z,z_{0}|q^{2},\omega \right) $ is the Fourier transform over in-plane
coordinate ${\bf R}_{\parallel }-{\bf R}_{0\parallel }$ and time of the
Green function $G$ from Eq.~(\ref{1}). This function obeys the equation

\begin{equation}
\left[ \frac{d^{2}}{dz^{2}}+q^{2}+i0+\delta \varepsilon \left( z\right) {\
\left( \frac{\omega }{c}\right) }^{2}\right] {\cal G}\left(
z,z_{0}|q^{2},\omega \right) =4\pi \delta \left( z-z_{0}\right) \ ,
\label{4}
\end{equation}

Formula (\ref{5}) for energy difference $\Delta q^{2}$ is valid, strictly
speaking, in the case of weakly dissipative medium, i.e. when
\begin{equation}
|\gamma (\omega )|\ll 1\ .  \label{6}
\end{equation}
Under the assumptions of weak dissipation and spectral narrowness of the
pulse the expression (\ref{5}) can be readily transformed to the form
\begin{equation}
\Delta q^{2}=\kappa _{2}^{2}-\kappa _{1}^{2}+2\varepsilon _{0}\frac{\omega
\Omega }{c^{2}}+i2\gamma (\omega )\left( \frac{\omega }{c}\right) ^{2}
\label{5a}
\end{equation}

Correlation functions of the type (\ref{3a}), (\ref{3}) with $\omega
_{1}=\omega _{2\text{ }}$(i.e. with $\Omega =0$) that appear in the theory
of stationary processes in 1D disordered systems can be calculated using
diagrammatic methods, \cite{21,15} functional method of Ref.~%
\onlinecite{AbrikRyzh}, or the resonance expansion method.\cite{6,IEEE} In
the next section the latter approach is shown to be quite universal and well
applicable (after some modification) also to non-stationary stochastic
problems, in particular, for calculation of the correlation function (\ref{3}%
) with $\Omega \neq 0$.


\section{Resonant scattering approximation for field correlators}


In this section, the method used in Refs.~\onlinecite{6,IEEE} for
calculating statistical moments of the field radiated by a monochromatic
point-like source is generalized to the case of pulse signals. The method
allows for rigorous calculation of the correlator (\ref{3}) provided a
single scattering can be regarded as weak.


\subsection{The resonance expansion method}

\label{subsec_3a} 

To calculate the intensity using Eq.~(\ref{2}) we have to know the Green
functions in (\ref{3}) for all values of $q^{2}$ in the interval $-\infty
<q^{2}<+\infty $. However, as it was shown in Ref.~\onlinecite{IEEE}, in the
case of weak scattering the contribution of spatial modes with $q^{2}<0$
(so-called evanescent modes) is significant only in a thickness of $%
|z-z_{0}|\sim \kappa _{m}^{-1}$ near the source position $z_{0}$. In the
rest of space the intensity is largely determined by the propagating
(extended) modes for which the Green function obeys equation (\ref{4}) with $%
q^{2}>0$. \ The key point of the following calculations is the so-called
resonance expansion of this Green function,

\begin{equation}
{\cal G}\left( z,z_{0}|q^{2},\omega \right) ={\cal G}_{1}(z,z_{0}){\rm e}%
^{iq(z-z_{0})}+{\cal G}_{2}(z,z_{0}){\rm e}^{-iq(z-z_{0})}+{\cal G}%
_{3}(z,z_{0}){\rm e}^{iq(z+z_{0})}+{\cal G}_{4}(z,z_{0}){\rm e}%
^{-iq(z+z_{0})}\ ,  \label{7}
\end{equation}
where ${\cal G}_{i}(z,z_{0})\equiv $ ${\cal G}_{i}\left(
z,z_{0}|q^{2},\omega \right) $ are smooth factors in comparison with the
``fast'' exponentials. The assumption of smoothness of the ``amplitudes'' $%
{\cal G}_{i}(z,z_{0})$ is based on the requirement for weak scattering (WS)
of the pulse-constituting plane harmonics, which means that the extinction
lengths $L$ of the harmonics, see Eqs.~(\ref{11}) below, are large compared
to their wavelengths and to the correlation radius $r_{c}$ of $\delta
\varepsilon (z)$ as well.

Formula (\ref{7}) represents the exact Green function as a sum of relatively
small packets of spatial harmonics centered at four basic ones, viz. $\exp
[\pm iq(z\pm z_{0})]$. Such a form of the solution of Eq.~(\ref{4} ) implies
that only {\em resonant} harmonics in the power spectrum of the permeability
fluctuation $\delta \varepsilon (z)$ contribute significantly to the
scattering of a wave with the wave number $q$, namely the harmonics with the
momenta close to zero, which are responsible for the forward scattering, and
close to $\pm 2q$ (backward scattering).

Using Green function in the form (\ref{7}) one can perform spatial averaging
of equation (\ref{4}) over an interval $2l$, such that $q^{-1}$, $r_{c}\ll
2l\ll L$. As the result, for the matrix

\begin{equation}
\hat{{\cal G}}=\left(
\begin{array}{cc}
{\cal G}_1 & {\cal G}_3 \\
{\cal G}_4 & {\cal G}_2
\end{array}
\right)  \label{8}
\end{equation}
of the smooth amplitudes from (\ref{7}) the equation follows

\begin{equation}
\left[ i\hat{\sigma}_{3}\frac{d}{dz}-\eta (z)-\hat{a}\zeta ^{\ast }(z)-\hat{a%
}^{\dagger }\zeta (z)\right] \hat{{\cal G}}(z,z_{0})=\frac{2\pi }{q}\delta
(z-z_{0})\ .  \label{9}
\end{equation}
Here $\hat{\sigma}_{3}$ and $\hat{a}$ are 2$\times $2 matrices
\[
\hat{\sigma}_{3}=\left(
\begin{array}{cc}
1 & \;\>0 \\
0 & -1
\end{array}
\right) \ ,\qquad \hat{a}=\left(
\begin{array}{cc}
0 & \;\>0 \\
1 & \;\>0
\end{array}
\right) \ ,
\]
superscript ($\dagger $) stands for Hermitian and the asterisk for complex
conjugation, respectively. Random functions (``potentials'') $\eta (z)$ and $%
\zeta (z)$ are constructed of narrow packets of spatial harmonics of $\delta
\varepsilon (z)$ as follows,
\begin{equation}
\eta (z)=\frac{-1}{2q}\left( \frac{\omega }{c}\right) ^{2}\int_{z-l}^{z+l}%
\frac{dz^{\prime }}{2l}\delta \varepsilon (z^{\prime })\ ,\qquad \zeta (z)=%
\frac{-1}{2q}\left( \frac{\omega }{c}\right) ^{2}\int_{z-l}^{z+l}\frac{%
dz^{\prime }}{2l}{\rm e}^{-2iqz^{\prime }}\delta \varepsilon (z^{\prime })\ .
\label{eta-zeta_def}
\end{equation}

On the assumption of weak scattering, functions $\eta (z)$ and $\zeta
(z)$ can be thought of as Gaussian random processes irrespective of
the statistics of $\delta \varepsilon (z)$. \cite{4} Correlation of
these functions was studied in detail in Ref.~\onlinecite{MakTar}
where the evidence was given that only two binary correlators of the
potentials (\ref {eta-zeta_def}), viz. $\left\langle \eta (z)\eta
(z^{\prime })\right\rangle $ and $\left\langle \zeta (z)\zeta ^{\ast
}(z^{\prime })\right\rangle $, are different from zero, and can be
replaced by weighted $\delta $-functions,

\begin{equation}
\left<\eta(z)\eta(z^{\prime})\right>=L_f^{-1}\delta(z-z^{\prime})\ ,\qquad
\left<\zeta(z)\zeta^*(z^{\prime})\right>=L_b^{-1}\delta(z-z^{\prime})\ .
\label{10}
\end{equation}
In (\ref{10}), length parameters $L_{f,b}$ are given by

\begin{equation}
L_{f}(q,\omega )=\left( {\frac{c}{\omega }}\right) ^{4}{\frac{(2q)^{2}}{%
\widetilde{W}(0)}}\ ,\qquad \qquad L_{b}(q,\omega )= \left( {\frac{c}{\omega}
}\right) ^{4}{\frac{(2q)^{2}}{\widetilde{W}(2q)}}\ ,  \label{11}
\end{equation}
$\widetilde{W}(p)$ is the Fourier transform of the binary correlation
function of the permeability fluctuations,

\begin{equation}
W\left( z-z^{\prime }\right) =\left\langle \delta \varepsilon (z)\delta
\varepsilon (z^{\prime })\right\rangle \ .  \label{12}
\end{equation}
It is shown in Ref.~\onlinecite{TarWRM} that (\ref{11}) are nothing but the
extinction lengths related to the forward ($f$) and backward ($b$)
scattering of the harmonics with the wave number $q$ and frequency $\omega $%
. In terms of these lengths, the WS conditions used when deriving equation (%
\ref{9}) are expressed through the inequalities
\begin{equation}
q^{-1},r_{c}\ll L_{f,b}\ .  \label{13}
\end{equation}
Notice that the relationship between small lengths $q^{-1}$ and $r_{c}$ is
of no crucial importance. It only specifies the Fourier components of the
correlation function (\ref{12}) and thus the possible distinction between
the ``forward'' and ``backward'' extinction lengths (\ref{11}).

For the resonance approximation (equivalent to WS requirement) to be most
efficient it is advantageous to represent both of the Green functions
entering the correlator (\ref{3}) in the form of the expansion (\ref{7})
with the same fast exponentials, i.e. with the same wave number $q$.
Although the amplitude functions ${\cal G}_{i}$ in (\ref{7}) cannot be found
explicitly, this representation proves to be quite helpful for the
calculation of the correlator (\ref{3}). Indeed, if we present both of Green
functions from (\ref{3}) in the form of expansion (\ref{7}) with the same
fast exponentials, only ``diagonal elements'' of the product $\hat{{\cal G}}%
\hat{{\cal G}}^{\ast }$ remain non-zero after the averaging (see next
subsection).

Note that functions ${\cal G}_{i}$ in (\ref{7}) can be recognized as slowly
varying amplitudes if along with the WS condition (\ref{13}) the inequality
holds 
\begin{equation}
\left| \Delta q^{2}\right| \ll q^{2}\ .  \label{14}
\end{equation}
Physically, this inequality is natural for the definition of weak scattering
since the quantity $q^{2}$ has the meaning of energy in Eq.~(\ref{4}). As it
will be clear from the subsequent calculation, the inequality (\ref{14}) is
coincident with the conditions (\ref{13}) supplemented by the requirements
of weak dissipation, $\gamma (\omega _{0})\ll 1$, and narrowness of the
pulse frequency band.

Green function of equation (\ref{4}) is the solution of a two-point
boundary-value problem with conditions given at $z\rightarrow \pm \infty $.
However, to systematically perform the averaging over random potentials
without resort to finite-order perturbation approximations (that fail to
take into account correctly the interference of multiply scattered waves in
one-dimensional random systems) it is much more convenient to deal with
random functions obeying Cauchy problems which are causal functionals of the
random potentials (\ref{eta-zeta_def}). Fortunately, the elements of Green
matrix (\ref{8}) can be factorized into products of the auxiliary
one-coordinate functions, each meeting the initial-value problem conditioned
at either plus or minus infinity. The factorization scheme is outlined in
Appendix~\ref{A}. Evolutional character of the equations for those functions
allows to obtain finite-difference equations (\ref{RnPn_eqs}) (see Refs.~%
\onlinecite{19,20}) for auxiliary correlators with the help of which the
correlation function (\ref{3}) can be appropriately calculated.


\subsection{Asymptotics of the correlation function Eq. (\ref{3})}

\label{subsec3b} 

To obtain analytic expressions for the correlation function (\ref{3}), we
assume the medium to be statistically uniform on average in $z$ direction,
and then pass from the coordinate representation of (\ref{3}) to its Fourier
transform over the variable $z-z_{0}$,
\begin{equation}
K(s)=\int_{-\infty }^{\infty }dz\>{\rm e}^{-is(z-z_{0})}K(z,z_{0})\ .
\label{20}
\end{equation}
After substitution of the matrix elements of Green matrix (\ref{8}) in the
form (\ref{G_1234}) into (\ref{3}), and then into (\ref{20}) (where all
non-coordinate arguments of function (\ref{3}) are omitted for a while), one
has to expand the factors $A^{\Delta }(z_{0})$ and $A^{\ast }(z_{0})$ of
those elements in series of the products $\Gamma _{+}^{\Delta }\Gamma
_{-}^{\Delta }$ and $\Gamma _{+}^{\ast }\Gamma _{-}^{\ast }$. Possibility of
such expansion is ensured by retro-attenuation $i0$ in Eq.~(\ref{4}). In the
course of statistical averaging, each of the terms of the double series
produced for the function $K(s)$ is decomposed into a product of functionals
of different causal types, viz. ``plus'' and ``minus'' type. Since the
potentials (\ref{eta-zeta_def}) are effectively $\delta $-correlated, the
supports of the random functions entering the functionals of different types
do not overlap. Therefore, statistical averaging of those functionals can be
performed independently.

It can be shown that the correlators of the type $\left\langle \left( \Gamma
_{\pm }^{\Delta }\right) ^{n}\left( \Gamma _{\pm }^{\ast }\right)
^{m}\right\rangle $ with $n\neq m$ in the double series for $K(s)$ are
exactly equal to zero. Indeed, from equations (\ref{pi-gamma}) it follows
that the functional series for the functions $\pi _{\pm }(z)$ consist solely
of the terms with equal numbers of the functional factors $\zeta $ and $%
\zeta ^{\ast }$, whereas the quantities $\gamma _{\pm }$ contain extra
factors, $\zeta ^{\ast }$ for $\gamma _{+}$ and $\zeta ^{\ast }$ for $\gamma
_{-}$. Inasmuch as under WS conditions (\ref{13}) functional variables (\ref
{eta-zeta_def}) can be regarded as Gaussian-distributed random fields, the
above-indicated correlators have non-zero values only if $n=m$.

The foregoing procedure has been described in detail in Refs.~%
\onlinecite{19,20}. Omitting here tedious calculations we present the final
result of the averaging. The function $K(s)$ is represented as a series,
\begin{equation}
K(s)=\left( \frac{2\pi }{q}\right) ^{2}\sum_{n=0}^{\infty }\left(
R_{n}+R_{n+1}\right) \left[ P_{n}(s)+P_{n}(-s)\right] \ ,  \label{21}
\end{equation}
where $R_{n}$ and $P_{n}(s)$ are the auxiliary correlation functions of the
form

\begin{subequations}
\label{RnPn}
\begin{equation}
R_{n}=\left\langle \left[ \Gamma _{\pm }^{\Delta }(z)\Gamma _{\pm }^{\ast
}(z)\right] ^{n}\right\rangle \ ,  \label{R_n}
\end{equation}
\begin{equation}
P_{n}(\pm s)=\pm \bigg<\left[ \Gamma _{\pm }^{\Delta }(z)\Gamma _{\pm
}^{\ast }(z)\right] ^{n}\int_{z}^{\pm \infty }\!\!\!dz^{\prime }\,\exp \big[%
is(z-z^{\prime })\big]\frac{\pi _{\pm }^{\Delta }(z^{\prime })\pi _{\pm
}^{\ast }(z^{\prime })+\gamma _{\pm }^{\Delta }(z^{\prime })\gamma _{\pm
}^{\ast }(z^{\prime })}{\pi _{\pm }^{\Delta }(z)\pi _{\pm }^{\ast }(z)}\bigg>%
\ ,  \label{P_n(s)}
\end{equation}
that obey the following finite-difference equations
\end{subequations}
\begin{subequations}
\label{RnPn_eqs}
\begin{equation}
\beta R_{n}-n\left( R_{n+1}+R_{n-1}-2R_{n}\right) +\alpha ^{2}\left( 1+2%
{\cal L}_{b}/{\cal L}_{f}\right) nR_{n}=0\ ,  \label{22}
\end{equation}
\begin{eqnarray}
- &&(n+1)^{2}\Big[P_{n+1}(s)-P_{n}(s)\Big]+n^{2}\Big[P_{n}(s)-P_{n-1}(s)\Big]%
+\left( is{\cal L}_{b}+\frac{\beta }{2}\right) P_{n}(s)+\beta nP_{n}(s)
\nonumber \\
+ &&\frac{\alpha ^{2}}{2}\Big[2n^{2}+2n+1+\frac{{\cal L}_{b}}{{\cal L}_{f}}%
(1+2n)^{2}\Big]P_{n}(s)={\cal L}_{b}\left( R_{n}+R_{n+1}\right) ,  \label{23}
\end{eqnarray}
where the following notations are used

\end{subequations}
\begin{equation}
\beta =-i\frac{\Delta q^{2}}{q}{\cal L}_{b}\ ,\qquad \alpha =\frac{(\omega
+\Omega )^{2}-\omega ^{2}}{\omega (\omega +\Omega )}\ ,\qquad {\cal L}%
_{f,b}=\left( \frac{\omega }{\omega +\Omega }\right) ^{2}L_{f,b}\ .
\label{24}
\end{equation}
Equations (\ref{RnPn_eqs}) have to be supplemented with the requirement that
functions $R_{n}$ and $P_{n}(s)$ tend to zero as $n\rightarrow \infty $. As
regards their behaviour at $n=0$, from definition (\ref{R_n}) it follows
that $R_{0}=1$ whereas for $P_{0}(s)$ integrability over the variable $s$ is
sufficient. \ Equations of this type have been studied in Refs.~%
\onlinecite{21,15,AbrikRyzh} and \onlinecite{19,20} in the context of the
conductivity of 1D disordered systems.

The terms proportional to $\alpha ^{2}$ allow, in principle, for arbitrary
non-stationarity of the wave to be taken into account. Yet narrowness of the
pulse frequency band assumed in this paper is consistent with the inequality
$|\alpha |\ll 1$ allowing for equations (\ref{RnPn_eqs}) to be solved
perturbatively in this parameter. In Appendix~B, it is demonstrated that if
the inequality holds $|\beta |\ll 1$, the summands $\propto \alpha ^{2}$ in
Eqs.~(\ref{RnPn_eqs}) contribute negligibly to the sum (\ref{21}), in
accordance with the condition (\ref{13}). As a consequence, in the limit of $%
|\beta |\rightarrow 0$ we arrive at the following expression for correlation
function (\ref{3}), 
\begin{equation}
K(z,z_{0};\kappa _{1},\kappa _{2};\omega ,\Omega )\approx \frac{i(2\pi )^{2}%
}{q{\cal L}_{b}\,\Delta q^{2}}\int_{0}^{\infty }d\mu \,W(\mu )\nu ^{2}(\mu
)\exp \left( -\frac{\nu (\mu )|z-z_{0}|}{{\cal L}_{b}}\right) \ .  \label{25}
\end{equation}
Here the notations are used
\begin{equation}
W(\mu )=\frac{\pi ^{2}}{2}\frac{\mu \sinh \left( \pi \mu /2\right) }{\cosh
^{3}\left( \pi \mu /2\right) }\ ,\qquad \nu (\mu )=\frac{1+\mu ^{2}}{4}\ .
\label{nu(mu)}
\end{equation}

In the limiting case $|\beta |\gg 1$, the terms proportional to $\alpha ^{2}$
in Eqs.~(\ref{RnPn_eqs}) result in small, though not {\em a priori}
negligible, corrections to the basic approximation for the correlation
function (\ref{3}), 
\begin{equation}
K(z,z_{0};\kappa _{1},\kappa _{2};\omega ,\Omega )\approx \left( \frac{2\pi
}{q}\right) ^{2}\left[ 1-\frac{\alpha ^{2}}{2}\left( 1+\frac{{\cal L}_{b}}{%
{\cal L}_{f}}\right) \frac{|z-z_{0}|}{{\cal L}_{b}}\right] \exp \left[
-\left( 1+\frac{\beta }{2}\right) \frac{|z-z_{0}|}{{\cal L}_{b}}\right] \ .
\label{26}
\end{equation}
It will be shown in the next section that the average intensity of a
narrow-band signal is mainly determined by the behaviour of the correlator (%
\ref{3}) at $\kappa _{1}\approx \kappa _{2}$, that corresponds to $|\beta
|\ll 1$ and, consequently, to the asymptotic expression (\ref{25}).


\section{Calculation of the pulse shape}

\label{shape} 

To calculate the average intensity $\left\langle I\left( {\bf R},{\bf R_{0}}%
|t\right) \right\rangle \ $we evaluate the integrals over $\Omega $ and $%
\omega $ in Eq.~(\ref{2}) with the correlator $K\left( z,z_{0}|\kappa
_{1},\kappa _{2};\omega ,\Omega \right) $ given by formula (\ref{25}) and
function $\varphi (\omega +\Omega -\omega _{0})$ presented in the form

\begin{equation}
\varphi (\omega +\Omega -\omega _{0})=\int_{-\infty }^{\infty }dt^{\prime
}A(t^{\prime })\,{\rm e}^{i(\omega +\Omega -\omega _{0})t^{\prime }}\ .
\label{27}
\end{equation}
From asymptotic expression (\ref{25}) it follows that in the lower
half-plane of the complex $\Omega $ their is a pole with ${\rm Im}\,\Omega
_{p}=-4\pi i\sigma /\varepsilon _{0}$, ($\Delta q=0)$ which allows to
calculate the integral over $\Omega $ as a residue in this point. Is is
obvious that a non-zero result is obtained for those values of $t^{\prime }$
that are limited by the condition
\begin{equation}
t^{\prime }<t^{\ast }(q_{2},\omega )=t-\tau (q_{2},\omega )\ .
\label{lim_t'}
\end{equation}
Here $\tilde{c}=c/\sqrt{\varepsilon _{0}}$, notation $\tau (q,\omega )$
stands for the time interval necessary for the plane harmonics $q$ to pass
from the plane of the source ($z_{0}$) to the plane of the receiver ($z$),
\begin{equation}
\tau (q,\omega )=\frac{|z-z_{0}|}{\tilde{c}q/\kappa _{m}}\ .  \label{def_tau}
\end{equation}

The next step is to calculate the integral over $\omega $ in Eq.~(\ref{2}).
Due to the presence of the narrow function $\varphi ^{\ast }(\omega -\omega
_{0}),$ all physical quantities in the integrand, in particular $\gamma
(\omega )$ and $\kappa _{m}(\omega )$, can be taken at the carrier frequency
$\omega _{0}$. Since in the present paper we are mainly interested in
localization effects, dissipation in the medium is supposed to be small
enough, and the dissipation rate of the carrier harmonics is much larger
than the corresponding extinction lengths (\ref{11}),
\begin{equation}
\frac{c}{\omega _{0}\gamma (\omega _{0})}\gg L_{f,b}(q,\omega _{0})\ .
\label{28}
\end{equation}
Subject to the condition (\ref{28}), the integral over $\omega $ recovers
the function $A^{\ast }(t^{\prime })$, whereupon the average intensity is
reduced to 
\begin{eqnarray}
\langle I  && \left({\bf R},{\bf R}_{0};t\right) \rangle \approx
\frac{{\tilde{c}}^{2}}{8\omega _{0}}\int \!\!\!\!\int_{0}^{\kappa _{m}^{2}}%
\frac{dq_{1}^{2}dq_{2}^{2}}{q_{2}L_{b}(q_{2})}\,J_{0}\left( \varrho \sqrt{%
\kappa _{m}^{2}-q_{1}^{2}}\right) J_{0}\left( \varrho \sqrt{\kappa
_{m}^{2}-q_{2}^{2}}\right)  \nonumber \\
&&\times \int_{-\infty }^{t^{\ast }(q_{2},\omega _{0})}dt^{\prime
}|A(t^{\prime })|^{2}\exp \left[ -\gamma \frac{\omega _{0}}{\varepsilon _{0}}%
(t-t^{\prime })+i\frac{{\tilde{c}}^{2}(t-t^{\prime })}{2\omega _{0}}%
(q_{1}^{2}-q_{2}^{2})\right]  
\int_{0}^{\infty }d\mu \,W(\mu )\nu ^{2}(\mu )\exp \left[ -\nu (\mu
)\frac{|z-z_{0}|}{L_{b}(q_{2})}\right] \ . \hspace{1cm} \label{30}
\end{eqnarray}

\vspace{.5\baselineskip}
From here on we address the case when the
receiver is located far from the source, so that
\begin{equation}
\kappa _{m}\varrho \gg 1\ .  \label{33}
\end{equation}
To integrate in Eq.~(\ref{30}) over $q_{1,2}$ we use the integral
representation of the Bessel functions and implement the saddle-point
method. Both of $q$-integrals in (\ref{30}) have the same saddle point
\begin{equation}
q_{s}^{2}=\kappa _{m}^{2}\left[ 1-\left( \frac{\varrho }{\tilde{c}%
(t-t^{\prime })}\right) ^{2}\right] \ ,  \label{saddle}
\end{equation}
so that simple calculation then yields
\begin{equation}
\left\langle I\left( {\bf R},{\bf R}_{0};t\right) \right\rangle \approx
\frac{\kappa _{m}^{2}}{2\omega _{0}} \int_{-\infty}^{t}\frac{dt^{\prime }}{%
(t-t^{\prime })^{2}}\frac{|A(t^{\prime })|^{2}}{q_{s}L_{b}(q_{s})} \exp %
\left[ -\gamma \frac{\omega _{0}}{\varepsilon _{0}}(t-t^{\prime })\right]
\int_{0}^{\infty }d\mu \,W(\mu )\nu ^{2}(\mu )\exp \left[ -\nu (\mu)%
\frac{|z-z_{0}|}{L_{b}(q_{s})}\right] \ .  \label{int_gen}
\end{equation}
This result for the space-time distribution of the average intensity
of a point-source-radiated narrow-band signal is rather general and
is valid for arbitrary envelope $A(t)$.  It is simplified
substantially when the distance $\varrho $ is large enough for the
pulse duration to be less than
the time of the pulse arrival at the observation point in homogeneous ($%
\delta \varepsilon (z)\equiv 0$) medium. In this case the upper limit
in the integral over $t^{\prime }$ in (\ref{30}) can be extended to
the infinity, all functions in the integrand of (\ref{int_gen}) can
be taken at $t^{\prime }=0$, and from (\ref{int_gen}) we obtain
\begin{equation}
\left\langle I\left( {\bf R},{\bf R}_{0};t\right) \right\rangle \approx
\frac{\kappa _{m}^{2}\tilde{c}\,T}{2(\omega _{0}t)^{2}L_{m}}\frac{\exp
\left( -\gamma \omega _{0}t/\varepsilon _{0}\right) }{\left[ 1-\left( {%
\varrho }/{\tilde{c}t}\right) ^{2}\right] ^{3/2}}\int_{0}^{\infty }d\mu
\,W(\mu )\nu ^{2}(\mu )\exp \left\{ -\frac{\nu (\mu )|z-z_{0}|}{L_{m}\left[
1-\left( {\varrho }/{\tilde{c}t}\right) ^{2}\right] }\right\} \ .  \label{35}
\end{equation}
Here $L_{m}=L_{b}(\kappa _{m},\omega _{0})$ is the largest value of the
backscattering-induced extinction length (localization length) corresponding
to the most ``energetical'' (i.e. $q_{2}=\kappa _{m}$) harmonics, and

\[
T=\int_{-\infty }^{\infty }dt^{\prime }\,|A(t^{\prime })|^{2}
\]

Although the intensity of a monochromatic field is known to be a strongly
fluctuating, not a self-averaged quantity in 1D disordered systems, the
integration of the correlator (\ref{25}) over parameters $\omega $ and $q$
(the last integration corresponds physically to the summation of plane
harmonics with different angles of propagation) serves as an additional
averaging factor that suppresses fluctuations of the intensity of the wave
packet, and therefore makes the results obtained by ansemble averaging, (\ref
{int_gen}) and (\ref{35}), more physically meaningful.


\section{Discussions of the results}


Equations (\ref{int_gen}) and (\ref{35}) present the space-time dependence
of the average intensity of a narrow-band pulse signal radiated by a point
source in a randomly layered weakly dissipative medium. Here we dwell on the
main physical characteristics of the result that are manifestations of the
strong Anderson localization in 1D disordered media.

First, it is evident from (\ref{int_gen}) and (\ref{35}) that the pulse
field is exponentially localized in $z$-direction within a $4L_{m}$ thick
layer whose central plane $z=z_{0}$ is the plane where the source is
located. In other words, the point-source pulse radiation is channelled,
much as the monochromatic radiation is, within the fluctuation waveguide
which is created owing to the interference of multiply back-scattered plane
harmonics even though the regular refraction does not exist in the system.

\begin{figure}[b!]
\setcaptionmargin{1cm} \captionstyle{hang} \centering
\vspace{.5\baselineskip}
\renewcommand{\captionlabeldelim}{: }
\scalebox{.7}[.6]{\includegraphics[bb=28 54 528 370]{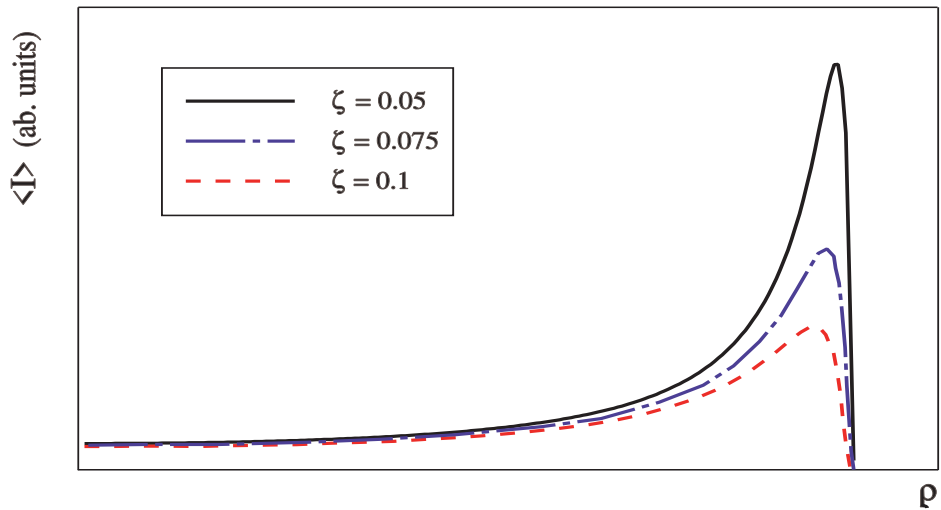}}
\caption{ Instantaneous spatial distributions of the intensity in the
fluctuation waveguide at different $\zeta =\frac{\left| z-z_{0}\right| }{%
4L_{m}}$.} \label{fig1}
\end{figure}

\begin{figure}[t!]
\setcaptionmargin{1cm} \captionstyle{hang}
\renewcommand{\captionlabeldelim}{: }
\centering \vspace{.8\baselineskip}
\scalebox{.7}[.65]{\includegraphics[bb=32 53 501 361]%
{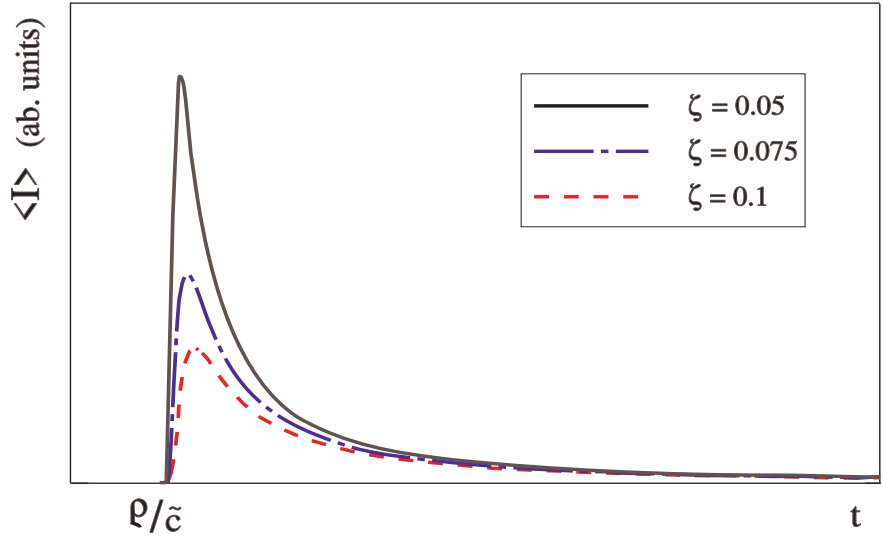}} \caption{Time dependences of the intensity at a given
in-plane distance $\varrho $ and different $\zeta =\frac{\left|
z-z_{0}\right|}{ 4L_{m}}$.} \label{fig2}
\end{figure}

The next interesting feature is that the narrow-band pulse acquires
universal shape at large distances in the fluctuation waveguide.
Indeed, under the assumption that in-plane distance to the
observation point is such that $\rho >\tilde{c}T$, the average
intensity is described by formula (\ref {35}) and depends on the
envelop of the incident pulse, $A(t),$ only through the normalization
constant $T$. In Fig.~\ref{fig1}, a set of curves is presented
depicting the intensity distribution as a function of the in-plane distance $%
\varrho $ at a given time $t$ for different distances in the
direction perpendicular to layers, $\zeta =\left| z-z_{0}\right|
/4L_{m}$. In Fig.~\ref{fig2}
the time dependence of the pulse intensity is shown at a certain distance $%
\varrho $ and three different $\zeta $. As it is seen from the
graphs, during the propagation in the fluctuation waveguide the
signal acquires a rather slowly decaying tail, and at large distances
from the maximum of the pulse the intensity decreases in time
proportionally to $t^{-2}$. The weak sensitivity of the wave packet
to its initial shape is due to the fact that in randomly layered
media (in distinction to free space) a point source radiates only
those eigen modes that are localized in a narrow (of the size of the
localization length) stripe near the source. \cite{FreiGred91,IEEE}
The (random) set of these modes is a fingerprint of each realization
of random potential and is independent on the way of excitation.

Another peculiarity of the pulse propagation in a randomly layered
medium is a sort of ``anisotropy'' of the time delay of the wave
packet: the arrival time increases with increasing $|z-z_{0}|$ faster
than it does when $\varrho$
grows. Indeed, the earliest signal arrival time at a point \{${\bf \varrho ,}%
z\}$ is of order $\frac{\sqrt{\varrho ^{2}+\left( z-z_{0}\right) ^{2}}}{{%
\tilde{c}}}$. At this moment, if $\varrho \gg L_{m}$, the spatial
distribution of pulse in the fluctuation waveguide ($|z-z_{0}|<L_m$)
given by Eq.~(\ref{35}) contains the exponentially small factor
\begin{floatingfigure}[r]{12cm}
\renewcommand{\captionlabeldelim}{: }
\centering  \vspace{.5\baselineskip} 
\scalebox{.7}[.7]{\includegraphics[bb=29 51 482 306]{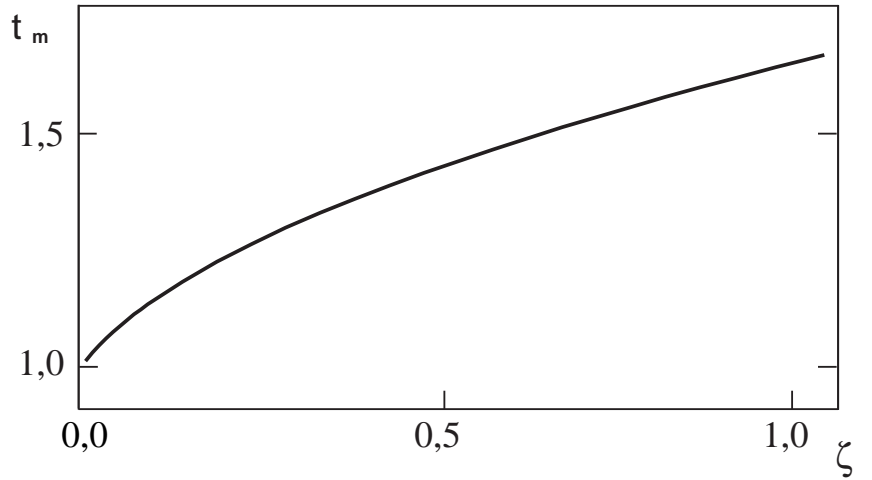}}
\caption{Arrival time of the pulse maximum, $t_{m}$ (in units
$\varrho /\tilde{c}$), vs $\zeta =\frac{\left| z-z_{0}\right|
}{4L_{m}}.$} \label{fig3}
\end{floatingfigure}

\begin{equation}
\exp \left[ -\nu (\mu )\frac{\varrho ^{2}}{L_{m}|z-z_{0}|\ }\right] \ll 1\ .
\end{equation}
The moment, $t_{m},$ when the signal at the point \{${\bf \varrho
,}z\}$ reaches its not exponentially small maximum can be roughly
estimated by equating the localization exponent in (\ref{35}) to
unity. Ths procedure yields

\begin{equation}
t_{m}\sim \frac{\varrho }{\tilde{c}}\left[ 1-\Xi \left( \frac{z-z_{0}}{L_{m}}%
\right) ^{2}\right] ^{-1/2}  \label{38}
\end{equation}
with some numerical coefficient $\Xi \sim 1$. Although the estimate (\ref{38}%
) cannot claim for satisfactory accuracy, it gives a reasonable idea
of the pulse delay in a stratified medium. To accurately calculate
the dependence of the arraival time of the pulse maximum as a fuction
of the transverse displacement $|z-z_{0}|$ one should use Eq.
(\ref{35}). The results of this (numerical) calculation is shown in
Fig.~\ref{fig3}.


\section{ Concluding remarks}


To summarize, in this paper the problem of the space-time
distribution of the average intensity of a narrow-band pulse which is
radiated by a point source in 3D randomly layered medium has been
solved by means of the generalized resonance expansion method. The
pulse field is shown to be localized in the direction of
stratification and channelled parallel to the layers within the
fluctuation waveguide whose symmetry plane goes through the source
location. The waveguide originates exclusively from the interference
of random fields multiply scattered by weak fluctuations, with no
regular refraction present in the medium. The typical width of the
waveguide is of order of the localization length of the harmonics
with largest allowable momentum along the axis of stratification.
Random lamination of the medium leads to a substantial distortion of
the pulse shape. Specifically, far away from the source the
narrow-band pulse of any original spectral content, being locked
within the fluctuation waveguide, spreads into a signal with the
envelope given by the universal function (\ref {35}) depicted in
Figs.~\ref{fig1} and \ref{fig2}. In contrast to homogeneous media,
the dependance of the arrival time of the pulse maximum on
coordinates is strongly anisotropic: it increases drastically as the
observation point moves in $z$-direction away from source. This delay
is not due to the increase of the path length of the signal, as it
is, for example, in media with regular refraction, but is caused by
the multiple random scattering of the saddle-point harmonics
(\ref{saddle}).

\appendix


\section{Calculation of the Green matrix Eq. (\ref{8})}
\label{A}

To find the 1D Green function of Eq.~(\ref{4}) we express it via solutions
of the appropriate Cauchy problems,
\begin{equation}
{\cal G}(z,z_{0})={\cal W}^{-1}\left[ \psi _{+}(z)\psi _{-}(z_{0})\theta
(z-z_{0})+\psi _{+}(z_{0})\psi _{-}(z)\theta (z_{0}-z)\right] \ .
\label{G-psi}
\end{equation}
Here, the functions $\psi _{\pm }(z)$ are the linear-independent solutions
of homogeneous equation (\ref{4}) with boundary conditions given at either
``plus'' or ``minus'' infinity, depending on the ``sign'' index, ${\cal W}$
is the Wronskian of those functions, $\theta (z)$ is the Heaviside unit-step
function.

In the case of real $q$, it is reasonable to represent the functions $\psi
_{\pm }(z)$ as superpositions of modulated harmonic waves propagating in
opposite directions of $z$-axis,
\begin{equation}
\psi _{\pm }(z)=\pi _{\pm }^{\Delta }(z)\exp (\pm iqz)-i\gamma _{\pm
}^{\Delta }(z)\exp (\mp iqz)\ .  \label{psi_pm}
\end{equation}
The upper index $\Delta $ indicates that the corresponding functions is
related to the first Green function in Eq.~(\ref{3}).

Under WS conditions, (\ref{13}), the ``amplitudes'' $\pi _{\pm }^{\Delta
}(z) $ and $\gamma _{\pm }^{\Delta }(z)$ in (\ref{psi_pm}) are smooth
functions in comparison with the nearby standing fast exponentials. By
averaging the equations for $\psi _{\pm }(z)$ over the interval of the
length $2l$ intermediate between ``small'' and ``large'' lengths of Eq.~(\ref
{13}) we arrive at a following set of dynamic equations,
\begin{subequations}
\label{pi-gamma}
\begin{eqnarray}
&&\pm {\pi _{\pm }^{\Delta }}^{\prime }(z)+i\big[\eta (z)-{\Delta q^{2}}/{2q}%
\big]\pi _{\pm }^{\Delta }(z)+\zeta _{\pm }(z)\gamma _{\pm }^{\Delta }(z)=0\
,  \label{pi-gamma-1} \\
&&\pm {\gamma _{\pm }^{\Delta }}^{\prime }(z)-i\big[\eta (z)-{\Delta q^{2}}/{%
2q}\big]\gamma _{\pm }^{\Delta }(z)+\zeta _{\pm }^{\ast }(z)\pi _{\pm
}^{\Delta }(z)=0\ .  \label{pi-gamma-2}
\end{eqnarray}
The function $\eta (z)$ in (\ref{pi-gamma}) coincides with the analogous
function from (\ref{eta-zeta_def}), with $\omega $ being replaced by $\omega
+\Omega $. The functions $\zeta _{\pm }(z)$ are given by
\end{subequations}
\begin{equation}
\zeta _{\pm }(z)=\frac{-1}{2q}\left( \frac{\omega +\Omega }{c}\right)
^{2}\int_{z-l}^{z+l}\frac{dz^{\prime }}{2l}{\rm e}^{\mp 2iqz^{\prime
}}\delta \varepsilon (z^{\prime })\ .  \label{zeta_pm}
\end{equation}
Sommerfeld's radiation conditions at $z\rightarrow \pm \infty $ are
reformulated as the ``initial'' conditions for the smooth amplitudes,
\begin{equation}
\lim_{z\rightarrow \pm \infty }\pi _{\pm }^{\Delta }(z)=1\ ,\qquad
\lim_{z\rightarrow \pm \infty }\gamma _{\pm }^{\Delta }(z)=0\ .
\label{in_cond}
\end{equation}
In a similar way the second Green function in (\ref{3}) is to be
represented, keeping in mind that in this case $\Delta q^{2}=0$ and $\Omega
=0$.

Wronskian ${\cal W}$ in (\ref{G-psi}) within the WS limit reduces to
\begin{equation}
{\cal W}=2iq\left[ \pi _{+}^{\Delta }(z)\pi _{-}^{\Delta }(z)+\gamma
_{+}^{\Delta }(z)\gamma _{-}^{\Delta }(z)\right] \ .  \label{Wronskian}
\end{equation}
By inserting then (\ref{psi_pm}) and (\ref{Wronskian}) into (\ref{G-psi})
and comparing the result with Eq.~(\ref{7}) we obtain for the matrix
elements of (\ref{8}):
\begin{subequations}
\label{G_1234}
\begin{eqnarray}
&&{\cal G}_{1,2}(z,z_{0}|\omega +\Omega ,q^{2}+\Delta q^{2})=-\frac{2\pi i}{q%
}A^{\Delta }(z_{0})\left[ \theta _{\pm }\frac{\pi _{\pm }^{\Delta }(z)}{\pi
_{\pm }^{\Delta }(z_{0})}-\theta _{\mp }\Gamma _{\pm }^{\Delta }(z_{0})\frac{%
\gamma _{\mp }^{\Delta }(z)}{\pi _{\mp }^{\Delta }(z_{0})}\right] \ ,
\label{G_12} \\
&&{\cal G}_{3,4}(z,z_{0}|\omega +\Omega ,q^{2}+\Delta q^{2})=\ -\frac{2\pi }{%
q}A^{\Delta }(z_{0})\left[ \theta _{\pm }\frac{\pi _{\pm }^{\Delta }(z)}{\pi
_{\pm }^{\Delta }(z_{0})}\Gamma _{\mp }^{\Delta }(z_{0})+\theta _{\mp }\frac{%
\gamma _{\mp }^{\Delta }(z)}{\pi _{\mp }^{\Delta }(z_{0})}\right] \ .
\label{G_34}
\end{eqnarray}
Here $\theta _{\pm }=\theta \big[\pm (z-z_{0})\big]$ and the rest of
notations are 
\end{subequations}
\begin{subequations}
\label{blocs_in_G}
\begin{eqnarray}
A^{\Delta }(z) &=&\left[ 1+\Gamma _{+}^{\Delta }(z)\Gamma _{-}^{\Delta }(z)%
\right] ^{-1}\ ,  \label{A(z)} \\
\Gamma _{\pm }^{\Delta }(z) &=&\gamma _{\pm }^{\Delta }(z)/\pi _{\pm
}^{\Delta }(z)\ .  \label{Ga_pm}
\end{eqnarray}
The upper sign indices in (\ref{G_1234}) correspond to ${\cal G}_{1}$ and $%
{\cal G}_{3}$ whereas the lower signs to ${\cal G}_{2}$ and ${\cal G}_{4}$,
respectively. The functions $\Gamma _{\pm }^{\Delta }(z)$ and $\pi _{\pm
}^{\Delta }(z)$ obey the Riccati-type coupled equations resulting directly
from (\ref{pi-gamma}),
\end{subequations}
\begin{subequations}
\label{Gamma_pi-eqs}
\begin{eqnarray}
&&\pm \frac{d\Gamma _{\pm }^{\Delta }(z)}{dz}=2i\left[ \eta (z)-\frac{\Delta
q^{2}}{2q}\right] \Gamma _{\pm }^{\Delta }(z)-\zeta _{\pm }^{\ast }(z)+\zeta
_{\pm }(z)\left[ \Gamma _{\pm }^{\Delta }(z)\right] ^{2}\ ,
\label{Gamma_pm-eq} \\
&&\pm \frac{d}{dz}\frac{1}{\pi _{\pm }^{\Delta }(z)}=i\left[ \eta (z)-\frac{%
\Delta q^{2}}{2q}\right] \frac{1}{\pi _{\pm }^{\Delta }(z)}+\zeta _{\pm }(z)%
\frac{\Gamma _{\pm }^{\Delta }(z)}{\pi _{\pm }^{\Delta }(z)}\ .
\label{1/pi_pm-eq}
\end{eqnarray}
The initial conditions for Eqs.~(\ref{Gamma_pi-eqs}) follow from (\ref
{in_cond}).

Transition from equations (\ref{pi-gamma}) to (\ref{Gamma_pi-eqs}) is
motivated by the following. In the stationary and non-dissipative limiting
case ($\Delta q^2=0$ and $\gamma=0$) the functions $\Gamma_\pm(z)$ represent
the amplitude reflection coefficients for the harmonics $q$ incident on the
1D disordered half-spaces ($z,\pm\infty$), respectively. In the presence of
arbitrary weak dissipation these functions become modulo less than unity
allowing for the factor $A^\Delta (z)$ given by Eq.~(\ref{A(z)}) to be
expanded into a series in powers of the product $\Gamma_+^\Delta\Gamma_-^%
\Delta$. Averaging then termwise the double series into which the product of
the Green functions in (\ref{3}) is expanded we arrive eventually at the
expression (\ref{21}).


\section{Analysis of the forward scattering contribution}


Equation (\ref{22}) can be solved rigorously at $\alpha =0$ (see,
e.g., Ref.~\onlinecite{21}), therefore it is not difficult to obtain
an asymptotic expression for $R_{n}$ at $|\alpha |\ll 1$. With the
accuracy of the first order in $\alpha ^{2}$ one can find that

\end{subequations}
\begin{equation}
R_n=\beta\int_{0}^{\infty}dt\,{\rm e}^{-\beta t} \left(\frac{t}{1+t}%
\right)^n \left\{ 1+\alpha^2\left(1+2\frac{{\cal L}_b}{{\cal L}_f}\right) t%
\left[1+t-\frac{\beta t}{6}(3+2t)\right]\right\}\ .  \label{B1}
\end{equation}
When $|\beta|\ll1$, the domain corresponding to $t\sim |\beta|^{-1}$ is
significant in the integral (\ref{B1}). Therefore the contribution of the
terms proportional to $\alpha^2$ is of the order $|{\alpha}/{\beta}|^2$. It
follows from Eqs.~(\ref{5}) and (\ref{24}) that

\begin{equation}
\left| \frac{\alpha }{\beta }\right| \lesssim \frac{\tilde c}{\omega {\cal L}%
_{b}}\ .  \label{B2}
\end{equation}
In rhs of (\ref{B2}) there is nothing but a small WS parameter which governs
all the approximations made in the course of solution. Therefore, the terms
proportional to $\alpha ^{2}$ in Eq.~(\ref{B1}) lead to the corrections that
are less than the calculation accuracy, and must be omitted in (\ref{B1}).

To analyze the equation (\ref{23}) we present the correlation function (\ref
{21}), using (\ref{B1}) at $\alpha =0$, in the integral form,

\begin{equation}
K(s)=\left(\frac{2\pi}{q}\right)^2\frac{{\rm e}^\beta}{\beta}
\int_\beta^\infty d\xi\,{\rm e}^{-\xi}(2\xi-\beta) \left[y_s(\xi)+y_{-s}(\xi)%
\right]\ .  \label{B3}
\end{equation}
Here $y_s(\xi)$ is the function related to the generating function $%
Y_s(\xi)=\sum_{n=0}^\infty z^n P_n(s)$ by the equality

\begin{equation}
y_s(\xi)=\frac{\xi}{\beta}Y_s\left(1-\frac{\xi}{\beta}\right) \ .  \label{B5}
\end{equation}
From Eq.~(\ref{23}) it follows that $y_s(\xi)$ obeys the differential
equation

\begin{eqnarray}\hspace{-1.5cm}
\Bigg[-\frac{d}{d\xi}\xi^2\frac{d}{d\xi} + \xi\frac{d}{d\xi}\xi +
\beta\frac{ d}{d\xi} \xi \left(\frac{d}{d\xi}- 1\right) + i\left( s
-\frac{\Delta q^2}{ 2q}\right){\cal L}_b\Bigg] && y_s(\xi) +
\frac{1}{2}\left(\frac{\alpha}{\beta}
\right)^2 \widehat Q_\beta y_s(\xi)  \nonumber \\
&&= {\cal L}_b\left[2+(2\xi-\beta){\rm e}^\xi {\rm Ei}(-\xi)+
\left(\frac{ \alpha}{\beta}\right)^2P_\beta (\xi)\right]\ ,
\label{B6}
\end{eqnarray}
where 
\begin{equation}
P_\beta(\xi)=\left(1+2\frac{{\cal L}_b}{{\cal L}_f}\right) {\rm e}%
^\beta\int_\beta^\infty d\xi^{\prime}{\rm e}^{-\xi^{\prime}} \frac{%
(2\xi^{\prime}-\beta)(\xi^{\prime}-\beta)}{\xi^{\prime}+(\xi-\beta)} \left[%
\xi^{\prime}-(\xi^{\prime}-\beta)(2\xi^{\prime}+\beta)\right] \ ,  \label{B7}
\end{equation}
and the differential operator $\widehat Q_\beta$ has the form
\begin{eqnarray}
\widehat Q_\beta =&& 2\left(\xi\frac{d}{d\xi}\xi\right)^2- \beta\left\{\xi%
\frac{d}{d\xi}\xi, \left\{\frac{d}{d\xi},\xi\right\}_+\right\}_+ +
\beta^2\left\{\xi\frac{d}{d\xi}\xi,\frac{d}{d\xi}\right\}_+ \cr\cr &&+ \frac{%
{\cal L}_b}{{\cal L}_f}\left[4\left(\xi\frac{d}{d\xi}\xi\right)^2+
4\beta\left(\xi\frac{d}{d\xi}\xi- \left\{\xi\frac{d}{d\xi}\xi,\frac{d}{d\xi}%
\xi\right\}_+\right)+ \beta^2\left(1-2\frac{d}{d\xi}\xi\right)^2\right]\ .
\label{B8}
\end{eqnarray}
The brackets $\{\cdots,\cdots\}_+$ in (\ref{B8}) denote an anticommutator.
It is evident from (\ref{B3}) that the solution of Eq.~(\ref{B6}) is of
importance in the domain $\xi\lesssim 1$. In that region the estimation is
valid

\[
\|\widehat Q_\beta\| \sim \left|P_\beta(\xi)\right| \sim 1+\frac{{\cal L}_b}{%
{\cal L}_f} \ .
\]
Thus we conclude that the terms proportional to $\widehat Q_\beta$
and $P_\beta(\xi)$, that contain the forward-scattering parameter
${\cal L}_f$, contribute negligibly, in accordance with WS parameter
(\ref{B2}), to the
solution of equation (\ref{B6}), just in the same way as the terms $%
\propto\alpha^2$ in (\ref{B1}).




\begin{thebibliography}{cc}
\bibitem{Isimaru}  A. Ishimaru, {\it Wave Propagation and Scattering in
Random Media}, (Academic, New York, 1978).

\bibitem{FreiGred91}  V. D. Freilikher and S. A. Gredeskul, Prog. Opt. {\bf %
30}, 137 (1991).

\bibitem{6}  Yu. V. Tarasov and V. D. Freilikher, Izv. Vyssh. Uchebn. Zaved.
Radiofiz (USSR) {\bf 32}, 1387 (1989) [SovRadiophys. (USA) {\bf 32}, 1024
(1989)]; Izv. Vyssh. Uchebn. Zaved. Radiofiz (USSR) {\bf 32}, 1494 (1989)
[Sov Radiophys. (USA) {\bf 32}, 1106 (1989)].

\bibitem{IEEE}  V. D. Freilikher and Yu. V. Tarasov, IEEE Trans. on AP {\bf %
39}, 197 (1991).

\bibitem{Wigner}  E. P. Wigner, Phys. Rev. {\bf 98}, 145 (1955).

\bibitem{21}  V. L. Berezinskii, Zh. Eksp. Teor. Fiz. (USSR) {\bf 65}, 1251
(1973) [Sov. Phys.---JETP {\bf 38}, 620, (1974)].

\bibitem{15}  A. A. Gogolin, Phys Rep {\bf 166}, 269 (1988).

\bibitem{AbrikRyzh}  A. A. Abrikosov and I. A. Ryzhkin, Adv. Phys. {\bf 27}
147 (1978).

\bibitem{4}  I. M. Lifshits, S. A. Gredeskul, and L. A. Pastur, {\it %
Introduction to the theory of disordered systems} (New York: Wiley, 1988)

\bibitem{MakTar}  N. M. Makarov and Yu. V. Tarasov, J. Phys.: Condens.
Matter {\bf 10}, 1523 (1998).

\bibitem{TarWRM}  Yu. V. Tarasov, Waves Random Media {\bf 10}, 395 (2000).

\bibitem{19}  E. A. Kaner and L. V. Chebotarev, Phys. Rep. {\bf 150}, 179
(1987).

\bibitem{20}  E. A. Kaner and Yu. V. Tarasov, Phys. Rep. {\bf 165}, 189
(1988).
\end{thebibliography}
\end{document}